\documentclass[12pt]{article}

\usepackage{amssymb}
\usepackage{amsmath}

\newtheorem{thrm}{Theorem}
\newtheorem{lem}{Lemma}
\newtheorem{prop}{Proposition}
\newtheorem{cor}{Corollary}

\newtheorem{defn}{Definition}
\newtheorem{exm}{Example}

\begin{document}

\begin{center}
{\bf\Large Partial actions and automata\footnote{This paper was
partially supported by CNPq and FAPESP (Brazil).}}
\end {center}

\begin{center}
{\bf M. Dokuchaev}\\ {\footnotesize Instituto de
Matem\'atica e Estat\'\i stica\\
Universidade de S\~ao Paulo, \\
Rua do Mat\~ao, 1010, CEP  05508-090, S\~ao Paulo, SP, Brazil
\\ E-mail: {\it dokucha@ime.usp.br}}
\end{center}

\begin{center}
{\bf B. Novikov}\\ { \footnotesize Kharkov National University,
Svobody sq., 4\\
61077, Kharkov, Ukraine
\\ E-mail: {\it boris.v.novikov@univer.kharkov.ua}}
\end{center}

\begin{center}
{\bf G. Zholtkevych}\\ { \footnotesize Kharkov National University,
Svobody sq., 4\\
61077, Kharkov, Ukraine
\\ E-mail: {\it g.zholtkevych@gmail.com}}
\end{center}

\begin{abstract}
We use the notion of a partial action of a monoid to introduce a
generalization of automata, which we call ``a preautomaton''. We
study properties of preautomata and of languages recognized by
preautomata.
\end{abstract}

\section*{Introduction}\label {sec-1}

The concept of the partial action has been introduced in \cite {exe}
for groups and extended to monoids in \cite {ms}. Therefore it is
natural to investigate the influence in Automata Theory of the
replacement of the full action of a free monoid by a partial action.
Our article is devoted to the study of this topic.

We propose the term ``preaction'' for a notion which is called
``partial action'' in \cite{ms} and ``strong partial action'' in
\cite{holl}. A generalization of the notion of an automaton which
appears here, will be called ``a preautomaton''. This change of the
terminology is caused by the fact that the term ``partial
automaton'' is widely used in Automata Theory in a different sense
(see, e.g., \cite{holc}).

We shall use the following facts from Automata Theory:

\begin{quote}
the Kleene theorem on regular languages,\\
the Myhill--Nerode theorem on languages and congruences,\\
the Eilenberg theorem on prefix decomposition.
\end{quote}
All of them can be found in \cite{eil,holc,hmu} etc.

\medskip

The paper begins with a preliminary section on preactions. Then we
define a preautomaton (Section \ref{sec-3}) and describe its
globalization. These notions are illustrated by an example in
Section \ref{sec-4}. Next we pass to languages which are recognized
by preautomata. The Eilenberg theorem is generalized and examples
are given which show that the Kleene theorem does not hold for
preautomata (Section \ref{sec-7}). In the last section we compare
the preautomata with other classes of machines.

We use the notation $ \varphi:A\dashrightarrow B $ for a partial
mapping from $A$ to $B$ (unlike to a full mapping $A\rightarrow B$).
If $\varphi (a)$ is not defined for $a\in A$ we write $\varphi (a) =
\emptyset$. The free semigroup on the alphabet $\Sigma$ is denoted
by $\Sigma^+$, the free monoid on $\Sigma$ by $\Sigma^\ast$ and its
identity element by $\varepsilon$. We use right preactions, as it is
customary in Automata Theory to deal with  right actions.

\section{Preactions}\label{sec-2}

We shall need some known information about partial actions
(preactions) of monoids which we recall in this section.

\begin{defn} \label{defn-1} {\rm \cite{ms}}
Let $M$ be a monoid with the identity $e$, $X$ a set. {\bf A
preaction} $M$ on $X$ is a partial mapping $X\times M
\dashrightarrow X:(x, a)\mapsto xa$, such that
$$
\forall x\in X \ \ xe=x,
$$
$$
\forall a, b\in M \ \ \forall x\in X \ \ xa\ne\emptyset\ \&\ (xa)
b\ne\emptyset\ \Longrightarrow\ x (ab) \ne\emptyset\ \&\ (xa) b=x
(ab),
$$
$$
\forall a, b\in M \ \ \forall x\in X \ \ xa\ne\emptyset\ \&\ x (ab)
\ne\emptyset\ \Longrightarrow\ (xa) b\ne\emptyset\ \&\ (xa) b=x
(ab).
$$
\end{defn}

The preactions of the given monoid $M$ form a category
$\mathcal{PA}\,M$: its objects are sets with preactions given on
them, and morphisms are such mappings $\varphi: X\to Y$ that
$$
\forall a\in M \ \forall x\in X \ \ xa\ne\emptyset\ \Longrightarrow\
\varphi (x) a\ne\emptyset\ \&\ \varphi (xa) = \varphi (x) a.
$$

Preactions appear in the following situation. Suppose that a full
action of a monoid $M$ is given on a set $Y$ and $X\subset Y$. Then
the restriction of the action to $X$ is a preaction.

Conversely, let $X\times M \dashrightarrow X$ be a preaction and
$Y\supset X$.

\begin{defn} \label{defn-2}
A full action $Y\times M \rightarrow Y$ is called a {\bf
globalization} if its restriction on $X$ coincides with the given
preaction.
\end{defn}

The following construction gives a globalization for any preaction
$X\times M \dashrightarrow X$. Define on the set $X\times M$ a
relation $\vdash$:
$$
(x, ab) \vdash (xa, b) \ \Longleftrightarrow\ xa\ne\emptyset.
$$

Let $\simeq$ is the equivalence generated by $\vdash$ and $Y =
(X\times M)/\simeq $. Denote by $[x,a]$ the $\simeq$-class
containing the pair $(x,a)$. Set $[x,a]b = [x,ab]$ for $[x,a]\in Y,\
b\in M$. This defines a full action on $Y$.

\begin{thrm} \label{thrm-1} {\rm \cite{ms}}
The above defined action $Y\times M \rightarrow Y $ is a
globalization of the preaction $X\times M \dashrightarrow X$; the
mapping $\alpha:X \rightarrow Y: x\mapsto [x,e]$ is an injective
morphism, and any morphism of the preaction $M$ on $X$ in a full
action of $M$ can be factored through $\alpha$.
\end{thrm}

\section{Preautomata}\label{sec-3}

We will use the definition of an automaton in the following form (in
this section the condition of finiteness is ignored):

\begin{defn}\label{defn-3}
Let $X$ be a set and $\Sigma ^\ast$ the free monoid over an alphabet
$\Sigma$. An {\bf automaton} is a triple $(\Sigma, X, \delta^\ast)$
where $\delta^\ast:X\times \Sigma^\ast \rightarrow X$ is a mapping
such that

\begin{equation}\label{eqn-1}
\forall x\in X \ \ \delta^\ast (x,\varepsilon)=x,
\end{equation}

\begin{equation}\label {eqn-2}
\forall u, v\in \Sigma ^\ast \ \forall x\in X \ \ \delta ^\ast (x,
uv) = \delta ^\ast (\delta ^\ast (x, u,), v).
\end{equation}
\end {defn}
The main object of this article is a more general concept:

\begin{defn}\label{defn-4}
A {\bf preautomaton} is such a triple $(\Sigma, X, \delta^\ast)$,
where $\delta^\ast:X\times \Sigma^\ast \dashrightarrow X$ is a
partial mapping, that

a) the above condition (\ref {eqn-1}) is satisfied;

b) if $\delta^\ast (x, u) \ne\emptyset$ and $\delta^\ast
(\delta^\ast (x, u), v) \ne\emptyset$, then $\delta^\ast (x, uv)
\ne\emptyset$ and
\begin{equation}\label{eqn-3}
\delta^\ast (x, uv) = \delta^\ast (\delta^\ast (x, u), v);
\end{equation}

c) if $\delta^\ast (x, u) \ne\emptyset$ and $\delta^\ast (x, uv)
\ne\emptyset$, then $\delta^\ast (\delta^\ast (x, u), v)
\ne\emptyset$ and also the equation (\ref{eqn-3}) is fulfilled.
\end{defn}

Clearly, preautomata correspond to preactions of the free monoid
$\Sigma^\ast$. It will be convenient to omit the symbol
$\delta^\ast$, i.\,e. to write $xu$ instead of $\delta^\ast (x,u)$,
and denote the preautomaton by $(\Sigma,X)$.

Theorem \ref{thrm-1} enables us to associate to the preautomaton
$\mathcal{M} = (\Sigma, X)$ an automaton $\mathcal{M}_{\rm gl} =
(\Sigma, Y)$, where $Y = (X\times \Sigma^\ast)/\simeq$ is the set
constructed in Section \ref{sec-2}. We call $\mathcal{M}_{\rm gl}$ a
{\bf globalization} of $\mathcal{M}$.

Using the fact that the monoid $\Sigma ^\ast$ is free, the
description of $\simeq $ can be simplified:

\begin{thrm}\label{thrm-2}
Define the following relation $\approx$ on the set $X\times
\Sigma^\ast$:
$$
(x,a) \approx (y,b) \Longleftrightarrow \exists \ a', b',p \in
\Sigma^\ast: a=a'p, \ b=b'p, \ xa'=yb'\ne\emptyset.
$$
Then $\approx$ coincides with $\simeq$.
\end{thrm}

{\bf Proof.}
If $(x, a) \approx (y, b)$ then $(x, a'p) \vdash (xa',p) = (yb',p)$
and $(y,b'p) \vdash (yb',p)$, whence $\approx\,\subseteq\,\simeq$.
Let us prove the converse inclusion.

Obviously, $\approx$ is reflexive and symmetric. Let us check its
transitivity.

Let $(x,a) \approx (y, b) \approx (z, d)$, $a=a'p$, $b=b'p=c'q$,
$d=d'q$ and $xa' =yb'\ne\emptyset$, $yc' =zd'\ne\emptyset$. Since
$\Sigma^\ast$ is free, either $p$ divides $q$ or vice versa. Let,
say, $p=uq$. Then $c' =b'u$.

Since $yb'\ne\emptyset$, $y(b'u) \ne\emptyset$ then $(xa')u =(yb')u
=y(b'u) \ne\emptyset$. Hence $(xa')u=x(a'u)$, as $xa'\ne\emptyset$.
On the other hand $a'p = (a'u)q$, and we obtain $(x, a'p) = (x,
(a'u)q) \approx (z,d'q)$, since $(xa')u = (yb')u=y(b'u) =yc' =zd'$.
Thus, $\approx$ is an equivalence, and $\approx\supset\simeq$ as
clearly $\approx\supset\vdash$.
$\blacksquare$

\begin{cor}\label{cor-1}
Every $\simeq$-class can be uniquely written either in the form $[x,
w]$, where $x\in X$, $w = a_1 \ldots a_n\in \Sigma^*$ and
$x(a_1\ldots a_i) = \emptyset$ for any $i, 1\le i\le n$, or in the
form $[x, \varepsilon]$, ($n=0$).
\end{cor}

Theorems \ref{thrm-1} and \ref{thrm-2} allow to give the following
interpretation of a finite preautomaton. Suppose that a system with
a possibly infinite set of states is given, and only a finite subset
of its states is accessible to our observation. Then the
preautomaton, obtained by restriction of the system to the set of
observable states, can be considered as a model of the observable
part of the system.

\section{Example}\label{sec-4}

Consider the infinite automaton $\mathcal{N} = (\Sigma, \mathbb{Z})$
for which $\Sigma = \{a, b\}$ is a two-lettered alphabet,
$\mathbb{Z}$ is the set of integers and the transition function is
defined by the equalities $n\cdot a=n+1$, $n\cdot b=n-1$. Observe
that this is the natural action of $\Sigma^\ast$ on its
factor-monoid by the congruence generated by the relation
$ab=\varepsilon$, and this factor-monoid is isomorphic to the
infinite cyclic group. Set $X = \{0,1\} \subset \mathbb{Z}$. Then
the restriction of the action of $\Sigma^*$ on $X$ gives a
preautomaton $\mathcal{M} = (\Sigma, X)$. In order to describe the
corresponding preaction, we denote the length of a word $w$ by
$|w|$, the number of entries of the letter $a$ in $w$ by $|w|_a$,
and set $\|w\| = |w|_a - |w|_b$. Then
$$
0\cdot w =\left \{
\begin{array}{ll}
  0 & \mbox{\rm if} \ \|w\| = 0, \\
  1 & \mbox{\rm if} \ \|w\| = 1, \\
  \emptyset & \mbox{\rm otherwise,}
\end{array}
 \right.
\ \ \ 1\cdot w =\left \{
\begin{array}{ll}
  0 & \mbox{\rm if} \ \|w\| =-1, \\
  1 & \mbox{\rm if} \ \|w\| = 0, \\
  \emptyset & \mbox{\rm otherwise.}
\end{array}
 \right.
$$

Corollary \ref{cor-1} allows to describe the globalization
$\mathcal{M}_{\rm gl} = (\Sigma, Y)$. A word
$w=a^{\alpha_1}b^{\beta_1}\ldots a^{\alpha_n} b^{\beta_n}$ will be
called {\it 1-simple} if $\|a^{\alpha_1}b^{\beta_1} \ldots
a^{\alpha_i}b^{\beta_i}\|>0$ for all $i\ge 1$ (this implies, in
particular, that $\alpha_1>0$). The word $\varepsilon$ will be
considered 1-simple too. Similarly, a word
$w=b^{\beta_1}a^{\alpha_1} \ldots b^{\beta_n}a^{\alpha_n}$ (and also
the word $\varepsilon$) is {\it 0-simple} if
$\|b^{\beta_1}a^{\alpha_1} \ldots b^{\beta_i}a^{\alpha_i}\|<0$ for
all $i\ge 1$. It is easy to see that every $\simeq$-class has  form
$[0, w]$ or $[1, w]$, where the word $w$ is 0-simple or 1-simple,
respectively.

Note that the preautomaton $\mathcal{M} = (\Sigma, X)$ cannot be
considered as a restriction of a finite automaton. Indeed, let
$\mathcal{P} = (\Sigma, W)$ be such an automaton with a transition
function $\delta:W\times\Sigma \to W$ and $X\subset W$. Since $W$ is
finite, $\delta (0, b^k) = \delta (0, b^m)$ for some distinct $k,m$.
Suppose $k<m$. As $\|b^ka^k\| = 0$ then $\delta (\delta (0, b^k),
a^k) = \delta (0, b^ka^k) =0\cdot (b^ka^k) =0$. Further, $\delta
(0,b^ma^k) = \delta (\delta (0, b^m), a^k) = \delta (\delta (0,b^k),
a^k) = 0$, and since our preautomaton is a restriction of
$\mathcal{P}$, then $0\cdot (b^ma^k) = 0$. This contradicts the
inequality $\|b^ma^k\| <0$.

In addition, it is possible to obtain from Theorem \ref{thrm-2} the
description of the semigroup of $\mathcal{M}_{\rm gl} = (\Sigma,Y)$.
We recall that the semigroup of an automaton is the factor-monoid
obtained from $\Sigma^*$ by identification of the words equally
acting of the states.

\begin{prop}\label{prop-1}
The semigroup of $\mathcal{M}_{\rm gl} = (\Sigma, Y)$ coincides with
$\Sigma^* = \{a, b\}^*$.
\end{prop}

{\bf Proof.}
Suppose that the words $u, v\in\Sigma^*$, $u\ne v$, act equally on
all states of $\mathcal{M}_{\rm gl} = (\Sigma, Y)$. In particular,
$[0, b^\beta]u = [0, b^\beta]v$ for all $\beta> 0$. As $b^\beta u\ne
b^\beta v$, then by Theorem \ref{thrm-2} $u=u'w$, $v=v'w$ and $0
\cdot b^\beta u' =0\cdot b^\beta v' \ne\emptyset$. Hence,

$$
\|u'\| = \|v'\| = \left\{
\begin {array} {l}
  \beta \\
  \beta+1
\end {array}
 \right.\ge \beta.
$$
But $\|u'\| = |u'|_a - |u'|_b\le |u'| \le |u|$ is a bounded quantity
in contrary with arbitrariness of $\beta$.
$\blacksquare$

\medskip

{\bf Remark.} It is easy to see that the semigroup of the original
automaton $\mathcal{N} = (\Sigma, Z)$ is isomorphic to $\mathbb{Z}$.

\section{Recognizability}\label{sec-5}

In what follows we will consider only {\bf finite} preautomata,
i.\,e. preautomata $\mathcal{M} = (\Sigma, X)$ such that $|X|
<\infty$. A preautomaton $\mathcal{M} = (\Sigma, X)$, in which {\bf
an initial state} $x_0\in X$ and {\bf a terminal subset} $T\subset
X$ are chosen, will be called {\bf a preacceptor} and denoted by
$\mathcal{M} = (\Sigma, X, x_0, T)$.

As well as in the classical situation, we call a language
$L\subset\Sigma^*$ {\bf recognizable} if there is a preacceptor
$\mathcal{M} = (\Sigma, X, x_0, T)$ for which
$$
L = \{w\in\Sigma^*\mid \emptyset\ne x_0w\in T \}.
$$
In what follows recognizability will be understood in this sense.

The following assertion generalizes the Myhill--Nerode theorem and
gives an algebraic characterization of recognizable languages:

\begin{thrm}\label{thrm-3}
A language $L\subset\Sigma^*$ is recognized by a preacceptor if and
only if $L$ is the union of a finite number of classes of some right
congruence on $\Sigma^*$.
\end{thrm}

{\bf Proof.}
Suppose that $L$ is recognized by a (finite) preacceptor
$\mathcal{M} = (\Sigma, X, x_0, T)$. Consider its globalization
$\mathcal{M}_{\rm gl} = (\Sigma, Y)$. It follows from Corollary \ref
{cor-1} that $L$ is recognized by the acceptor
$$
\mathcal{N} = (\Sigma, [x_0, \varepsilon]\Sigma^*, [x_0,
\varepsilon], [T, \varepsilon])
$$
where $[T, \varepsilon] = \{[t,
\varepsilon] \mid t\in T \}$, $[x_0, \varepsilon]\Sigma^*\subseteq
Y.$ The relation
$$
\rho=\{(u,v)\in\Sigma^*\times\Sigma^*\mid [x_0, \varepsilon]u =
[x_0, \varepsilon]v\}
$$
is a right congruence, $L$ is a union of $\rho $-classes and the
number of these classes does not exceed $|[T, \varepsilon]| = |T|
<\infty$.

Conversely, let $\rho$ be a right congruence on $\Sigma^*$ and $L$
be the union of a finite number (say, $n$) of $\rho$-classes $C_1,
\ldots, C_n$.

In the (infinite) automaton $\mathcal{K} = (\Sigma, \Sigma^*/\rho)$
we choose the $\rho$-class $E$ containing $\varepsilon$ as the
initial state, and $T = \{C_1, \ldots, C_n \} \subset \Sigma^*/\rho$
as the terminal subset. Then $(\Sigma, T\cup \{E\}, E, T)$ is a
finite preacceptor which is a restriction of $\mathcal{K}$ and
recognizes $L$.
$\blacksquare$

\begin{cor}\label{cor-2}
Given languages $L, M\subset\Sigma^*$ which are recognized by
preacceptors, then $L\cap M$ is also recognizable.
\end{cor}

{\bf Proof.}
Let $L$ and $M$ are finite unions of classes of right congruences
$\lambda$ and $\mu$ respectively. Then $L\cap M$ is a union of a
finite number of classes of the congruence $\lambda\cap \mu$.
$\blacksquare$

\begin{cor}\label{cor-3}
For the single-letter alphabet $\Sigma = \{a\}$ a language
$L\subset\Sigma^*$ is recognizable by a preacceptor exactly when $L$
is recognizable by an acceptor.
\end{cor}

{\bf Proof.}
The semigroup $\Sigma^+ = \{a\}^+$  is the unique infinite monogenic
semigroup up to isomorphism \cite {la}. As it is commutative, all
its right congruences are two-sided ones. Factor-semigroups by these
congruences (except the trivial one) are finite. Therefore if a
language $L$ is recognized by a preacceptor, but not by an acceptor,
it should be a union of a finite number of classes of the trivial
congruence, i.\,e. $L$ is  a finite subset in $\Sigma^+$, hence $L$
is regular.
$\blacksquare$

\begin{exm}\label{exm-1}
\rm The language $L = \{a^nb^n\mid n> 0 \}$ over the alphabet
$\Sigma=\{a, b\}$ is a class of the right syntactic congruence
\cite{la}. Therefore it is recognized by a preacceptor. As it is
well-known \cite{la}, $L$ is not recognized by any finite acceptor.
\end{exm}

\begin{exm}\label{exm-2}
\rm For an arbitrary finite $\Sigma$ each ideal of the monoid
$\Sigma^*$ is recognizable, since it is an element of the Rees
factor-semigroup.
\end{exm}

\section{Minimization of a preacceptor}\label{sec-6}

The study of recognizability of languages by preacceptors leads to
the notion of the syntactic equivalence of preacceptors.

\begin{defn}\label{defn-5}
Preacceptors $\mathcal{M}_1$ and $\mathcal{M}_2$ over the same
alphabet $\Sigma$ are called {\bf syntactically equivalent} if they
recognize the same language.
\end{defn}

As in the theory of acceptors, a question arises how to find in a
class of syntactically equivalent preacceptors a preacceptor whose
set of states has minimal cardinality. The first step is given by
the following lemma.

\begin{lem}\label{lem-1}
Let $\mathcal{M} = (\Sigma, X, x_0, T)$ be a preacceptor, $Y =
\{x_0\} \cup T$, $\mathcal{M}_0 = (\Sigma, Y, x_0, T)$ the
restriction of $\mathcal{M}$ on $Y$. Then $\mathcal{M}_0$ is
syntactically equivalent to $\mathcal{M}$.
\end{lem}

{\bf Proof.}
is obvious.
$\blacksquare$

\medskip

Thanks to Lemma \ref{lem-1}, in the study of language recognition of
preacceptors we can restrict the set of states including only the
initial and terminal states.

We recall from \cite{la} that the right syntactic congruence on
$\Sigma^*$ of a language $L \subset \Sigma^*$ is the relation
$\equiv_{L}$ defined by
$$
w_1 \equiv_{L} w_2 \Longleftrightarrow \forall {u \in \Sigma^*} \
(w_1u \in L \Leftrightarrow w_2u \in L)
$$
The language $L$ is the union of some classes of this relation;
moreover, any right congruence, such that $L$ is a union of its
classes, is contained in $\equiv_{L}$. This property allows us to
reformulate Theorem \ref{thrm-3} as follows:

\begin{lem}\label{lem-11}
A language $L\subset\Sigma^*$ is recognized by a (finite)
preacceptor if and only if it is the union of a finite number of
classes of its right syntactic congruence.
\end{lem}

We will denote the $\equiv_L$-class containing $u\in\Sigma^*$ by
$[u]_{\equiv_L}$. Fix a set of representatives $u_1, \dots, u_k\in
L$ of the $\equiv_L$-classes of $L$.

For the set $X_{\equiv_L} = \{[\varepsilon]_{\equiv_L},
[u_1]_{\equiv_L}, \dots, [u_k]_{\equiv_L} \} \subset
\Sigma^*/\equiv_L$ we define the partially defined map $\delta^*
\colon X_{\equiv_L} \times\Sigma ^*\dashrightarrow X_{\equiv_L}$ as
follows:

\begin{equation}\label{eqn-19}
\forall \, [u]_{\equiv_L} \in X_{\equiv_L}\ \ \ \ \delta^*
([u]_{\equiv_L}, \varepsilon) = [u]_{\equiv_L},
\end{equation}
\begin{equation}\label{eqn-20}
\forall \, [u]_{\equiv_L} \in X_{\equiv_L} \ \forall \,w\in\Sigma^+
\ \ \ uw\notin L \Rightarrow \delta^* ([u]_{\equiv_L}, w) =
\emptyset,
\end{equation}
\begin{equation}\label{eqn-21}
\forall \, [u]_{\equiv_L} \in X_{\equiv_L}\ \ \forall \, w\in\Sigma^
+ \ \  \ uw\in L \Rightarrow \delta^* ([u]_{\equiv_L}, w) =
[uw]_{\equiv_L}.
\end{equation}
\begin{thrm}\label{thrm-10}
Let $L\subset\Sigma^*$ is a finite union of $\equiv_L$-classes. Then
the preacceptor
$$
\mathcal{M}_{\equiv_L} = (\Sigma, X_{\equiv_L},
[\varepsilon]_{\equiv_L}, \{[u]_{\equiv_L} \mid u\in L \})
$$
with the partial mapping $\delta^*$, defined by
(\ref{eqn-19})--(\ref{eqn-21}), recognizes $L$.

Moreover, $\mathcal{M}_{\equiv_L}$ has the minimal set of states
among all finite preacceptors recognizing $L$.
\end{thrm}

{\bf Proof.}
First of all we check that $\delta^*$ defines a preaction. Indeed,
condition a) of Definition \ref{defn-4} is fulfilled by virtue of
(\ref{eqn-19}), and conditions b) and c) follow from (\ref{eqn-20})
and (\ref{eqn-21}). Thus, $\mathcal{M}_{\equiv_L}$ is a preacceptor
which recognizes $L$.

Now, let $\mathcal{M} = (\Sigma, X, x_0, T)$ be a preacceptor
recognizing $L$. Define a mapping $f\colon X_{\equiv_L} \to X$ by
the formulas
$$
f([\varepsilon]_{\equiv_L}) =x_0,
$$
$$
f([u_i]_{\equiv_L}) =x_0\cdot u_i, \ \ i=1, \dots, k.
$$
As $u_i\in L$, $x_0\cdot u_i\in T$.

We wish to show that $f$ is injective. Assume that $x_0\cdot
u_i=x_0\cdot u_j$ for some $1 \le i\neq j\le k$. Suppose that
$u_iw\in L$, so $\emptyset\neq x_0\cdot (u_iw) \in T$. Since $u_i\in
L$, then $\emptyset\neq (x_0\cdot u_i) \cdot w=x_0\cdot (u_iw) \in
T$ by the definition of a  preautomaton. The assumption $x_0\cdot
u_i=x_0\cdot u_j$ implies $\emptyset\neq x_0\cdot (u_jw) = (x_0\cdot
u_j) \cdot w\in T$, i.\,e. $u_jw\in L$. Thus $u_iw\in L$ yields
$u_jw\in L$, and similarly $u_jw\in L \Longrightarrow u_iw\in L$.
Hence $u_i\equiv_Lu_j$, contradicting $i\ne j$. Thus $f$ is
injective and $|X_{\equiv_L}| \le |X|$.
$\blacksquare$

\section{Prefix decomposition}\label{sec-7}

In this section Theorem VI.4.1 from \cite{eil} is generalized to
finite preautomata.

\begin{lem}\label{lem-4}
Suppose that $L$ is a language over an alphabet $\Sigma$ which is
recognized by a preacceptor $\mathcal{M} = (\Sigma, X, x_0, T)$ and
let $L_y$ ($y\in T$) be languages such that $L_y$ is recognized by a
preacceptor $\mathcal{M}_y = (\Sigma, X, x_0, T_y = \{y \})$. Then
$\displaystyle L =\coprod_{y\in T} L_y$ (disjoint union).
\end{lem}

{\bf Proof.}
$w\in L$ if and only if $x_0w=y$ for some $y\in T$, i.\,e. $w\in
L_y$.
$\blacksquare$

\medskip

We remind that a (possibly, empty) language $L\subset\Sigma^\ast$ is
called {\bf a prefix code} \cite{la} if $u, uv\in L $ implies
$v=\varepsilon$. The language $\{\varepsilon \}$ is also considered
as a prefix code (note that if a prefix code contains $\varepsilon$
then it is equal $\{\varepsilon \}$). It is well-known and easily
verified that the monoid $S=L^\ast$, generated by a prefix code $L$,
is free. Such a monoid is called {\bf unitary} and is characterized
by the property:
\begin{equation}\label{eqn-4}
u, uv\in S\Longrightarrow v\in S.
\end{equation}

The following lemma generalizes (\ref{eqn-4}):

\begin{lem}\label{lem-5}
Let $H$ and $C$ be prefix codes over an alphabet $\Sigma $. Then
$$
u, uv\in HC^\ast\Longrightarrow v\in C^\ast.
$$
\end{lem}
{\bf Proof.}
Suppose that
$$
u=hc_1\ldots c_m, \ uv=h'c'_1\ldots c'_n \ (h, h'\in H, c_i, c'_j\in
C).
$$
Then $hc_1\ldots c_mv=h'c'_1\ldots c'_n$. Since $H$ is a prefix
code, then $h=h'$, and hence $c_1\ldots c_mv=c'_1\ldots c'_n$. This
implies $c_i = c'_i$ $ (1\le i\le m\le n)$ and $v=c'_{m+1} \ldots
c'_n\in C^\ast$.
$\blacksquare$

\medskip

It is interesting to note that the property of prefix codes to
generate free monoids can be generalized as follows:

\begin{prop}\label{lem-6}\footnote{Proposition \ref{lem-6} will not be used in this article.}
Let $P_1, \ldots, P_n \ (n\ge 1)$ be prefix codes over $\Sigma$ and
$w\in P_1 \cdot\ldots\cdot P_n$. Then the decomposition $w=w_1\ldots
w_n$, where $w_i\in P_i$ for $i = 1, \ldots, n$, is unique.
\end{prop}

{\bf Proof.}
We use induction on $n$. For $n=1$ the statement is evident. Suppose
that it is true for $k <n$. Let $w=w'_1\ldots w'_n=w''_1\ldots
w''_n$ be two representations of a word $w\in P_1 \cdot\ldots\cdot
P_n $ where $w'_i, w''_i\in P_i$ for $i = 1, \ldots, n$. Denote $u'
=w'_2\ldots w'_n$, $u''= w''_2\ldots w''_n$ and assume that $|u'|
\ne |u''|$, for example, $|u'| <|u''|$. Then $|w'_1| >|w''_1|$, so
$w''_1$ is a prefix of $w'_1$, which is impossible for a prefix
code.

Hence, $|u'| = |u''|$. Then $u' =u''$ as suffixes of $w$ of the same
length, and $w'_1=w''_1$ as prefixes. By induction the equality $u'
=u''$ implies $w'_2=w''_2, \ldots, w'_n=w''_n$.
$\blacksquare$

\begin{defn} \label{defn-41}
A language $L\in\Sigma^\ast$, such that $L=HC^\ast$ where $H$ and
$C$ are prefix codes over $\Sigma$, will be called a {\bf
$p$-language}.
\end{defn}

{\bf Remark.} Notice that if $H=\emptyset$ then $HC^\ast=\emptyset$.

\begin{thrm}\label{thrm-4}
A language $L$ over an alphabet $\Sigma$ is recognized by some
preacceptor $\mathcal{M} = (\Sigma, X, x_0, \{y\})$ with a single
terminal state if and only if $L$ is a $p$-language.
\end{thrm}

{\bf Proof.}
Let $L$ be a $p$-language. We consider the possible cases.

1) $L_1 =\emptyset$. Then $L_1$ is recognized by the preacceptor
$$
\mathcal{M} = (\Sigma, \{x_0, y\}, \{x_0\}, \{y\})
$$
where the action is trivial:
$$
x_0w=x_0, \ yw=y.
$$

2) $L_2=C^\ast.$ Then $L_2$ is recognized by the preacceptor
$$
\mathcal {M} = (\Sigma, \{x_0\}, \{x_0\}, \{x_0\})
$$
where preaction It is given by formulas:
$$
x_0w =\left\{
\begin{array} {ll}
  x_0 & \mbox{\rm if} \ w\in C^\ast, \\
  \emptyset & \mbox{\rm if} \ w\notin C^\ast.
\end{array}
 \right.
$$
The fact that the above is a preaction, is verified using
(\ref{eqn-4}).

\medskip

3) $L_3=HC^\ast$ with $H\ne \{\varepsilon\}$. Then $L_3$ is
recognized by the preacceptor $\mathcal{M} = (\Sigma, \{x_0, y\},
\{x_0\}, \{y\})$ with the preaction:
$$
x_0w =\left\{
\begin{array}{ll}
  y, & \mbox {\rm if} \ w\in HC^\ast, \\
  x_0, & \mbox {\rm if} \ w =\varepsilon, \\
\emptyset, & \mbox {\rm if} \ w\not\in HC^\ast\cup \{\varepsilon \},
\end{array}
 \right.
\ \ \ yw =\left\{
\begin{array}{ll}
  y & \mbox {\rm if} \ w\in C^\ast, \\
\emptyset & \mbox {\rm if} \ w\not\in C^\ast.
\end{array}
 \right.
$$

We need to show that the above formulas define a preautomaton. For
the action of an input word on the state $y$ the fulfilment of the
conditions of Definition \ref{defn-4} follows from (\ref{eqn-4}). As
to the action on $x_0$, the condition b) of Definition \ref{defn-4}
is immediate, and in order to see c) suppose that $x_0u\ne\emptyset$
and $x_0 (uv) \ne\emptyset$. If $u =\varepsilon$, the condition c)
is obvious. Otherwise $u,uv\in HC^\ast$ and by Lemma \ref{lem-5},
$v\in C^\ast$.

Now we prove that if a language $L$ is recognized by a preacceptor
of the form $\mathcal{M} = (\Sigma, X, x_0, \{y\})$ then it is a
$p$-language. For every word $w\in\Sigma^\ast$ denote by ${\rm
Px}(w)$ the set of all proper prefixes of $w$. For
$K\subset\Sigma^\ast$ write ${\rm Px}(K) = \{{\rm Px} (w) \mid w\in
K\}$. Set
$$
H=\{w\in\Sigma^\ast\mid x_0w=y\ \&\ \forall u\in {\rm Px}(w) \
x_0u\ne y\}.
$$
and similarly,
$$
C=\{w\in\Sigma^\ast\mid yw=y\ \&\ \forall u\in {\rm Px}(w)\ yu\ne
y\}.
$$
By construction $H$ and $C$ are prefix codes and, moreover,
$L\supset HC ^\ast$ since $x_0w=y$ for any word $w\in HC^\ast$. It
remains to show that $L\subset HC^\ast $.

Let $w\in L$, i.\,e. $x_0w=y$. First assume that $x_0=y$. Write
$w=uv $, where $u$ is the least prefix of $w$ for which $x_0u=x_0$
(and therefore, $u\in H=C$). By definition of a preaction
$\emptyset\ne (x_0u)v=x_0w$, whence $x_0v=x_0$, and consequently
$v\in C^\ast$. Thus we obtain $w=uv\in HC^\ast=C^\ast$.

If $x_0\ne y$, then write $w=uv$, where $u$ is the prefix of the
least length, such that $x_0u=y$. It follows that $u\in H$ and
$yv=y$. As above, $v\in C^\ast$, whence $w\in HC^\ast$.
$\blacksquare$

Theorem \ref{thrm-4} allows us to transfer to preautomata the
concept of a prefix decomposition (\cite{eil}, Theorem VI.4.1):

\begin{cor}\label{cor-4}
If a language $L$ over the alphabet $\Sigma$ is recognized by some
preacceptor then $L$ decomposes into a disjoint union
$$
L =\coprod_iH_iC_i^\ast,
$$
where $H_i, C_i$ are prefix codes over $\Sigma$.
\end{cor}

{\bf Proof.}
The corollary follows from Theorem \ref{thrm-4} and Lemma
\ref{lem-4}.
$\blacksquare$

\medskip

The converse statement does not hold:

\begin{exm}\label{exm-3}
\rm It is directly seen that the language $H_1 = \{a\}^+$ is
recognized (even by a finite acceptor), and the language
$H_2=\{a^nb^n \mid n\ge 1\}$ is a prefix code. Take
$C_1=C_2=\{\varepsilon\}$. The union $H=H_1\cup H_2=H_1C_1^\ast\cup
H_2C_2^\ast$ is not recognized by any preacceptor. Indeed, let
$\equiv_H$ be the right syntactic congruence of $H$, and take $x\in
\Sigma^\ast$. Since
$$
a^nx\in H\Longleftrightarrow x=b^n\ \mbox {or} \ x\in H_1 \cup
\{\varepsilon \},
$$
the words $a^n$ form one-element $\equiv_H$-classes, and by Theorem
\ref{thrm-3} the language $H$ is non-recognizable.
\end{exm}

The given example shows also that the set of languages, recognized
by finite preacceptors, is not union-closed. Moreover, it is not
closed with respect to the  other Kleene's operations, i.e. the
product and the iteration:

\begin{exm}\label{exm-4} \rm Let $H_1$
and $H_2$ be the same as in Example \ref{exm-3}. By Theorem
\ref{thrm-4} the language $H_2^\ast$ is recognizable. Consider the
product $K=H_2^\ast H_1$. Then $a^n\in K$ for $n>0$ and
$$
\{x\in \Sigma^\ast \mid a^nx\in K\}\cap\{b\}^+=\{b^n\}.
$$
Therefore all words of the form $a^n$ are pairwise non-equivalent
with respect to the right congruence $\equiv_K$. As in Example
\ref{exm-3}, it follows that $K$ is non-recognized.
\end{exm}

\begin{exm}\label{exm-5}
\rm The language $L=H_2\cup \{a \}$ is recognized by a preacceptor,
since it consists of two $\equiv_L$-classes $H_2$ and $\{a\}$.
Consider the iteration $M=L^\ast$. It is easy to see that $a^mb^n\in
M$ iff $m\ge n$. Hence,
$$
\{x\in\Sigma^\ast \mid a^mx\in M\}\cap\{b\}^+=\{b^k\mid k\le m\}.
$$
As above, the words $a^m$ are pairwise non-equivalent with respect
to $\equiv_M$. Therefore, $M=L^\ast$ is not recognized by any
preacceptor.
\end{exm}

\section*{Conclusion}\label{sec-8}


The obtained results allow to clarify relations between the class of
finite preautomata \textsf{FPA} and other types of machines.

Obviously, \textsf{FPA} includes the class of finite automata
\textsf{FA}. This inclusion is strict since the language from
Example \ref{exm-1} is not recognized by any finite automaton. On
the other hand, it is known \cite[\S\,9.1]{la} that the prefix
language $\{a^nb^nc^n \mid n\ge 1\}$ is not context-free. According
to Theorem \ref{thrm-4} it means that \textsf{FPA} is not contained
in the class of automata with stack memory \textsf{FSA}.

At the same time, \textsf{FPA} does not include the class of Turing
machines \textsf{TM}. Indeed, the language $\{a^{n^2} \mid n\ge 1\}$
is not recognized by a finite automaton \cite[ex.\,4.1.2]{hmu}, and
by Corollary \ref{cor-3} it is not recognized also by a
preautomaton.

On the other hand,  the cardinal of the class of all prefix codes
over some finite alphabet equals to the cardinal of continuum,
whereas the class of a recursively enumerable languages over a
finite alphabet is countable. It follows that \textsf{TM} does not
contain \textsf{FPA}.

\end{document}